\documentclass[10pt]{article}
\usepackage{amsmath}
\usepackage{graphicx}
\usepackage{amsfonts}
\usepackage{amssymb}
\usepackage{epsf}
\usepackage{latexsym}

\def\be{\begin{equation}}
\def\ee{\end{equation}}
\def\bea{\begin{eqnarray}}
\def\eea{\end{eqnarray}}
\def\pa{\partial}

\def\fn{\footnote}

\def\B{\mbox{\tiny B}}
\def\d{\textrm{d}}

\def\case#1/#2{\textstyle\frac{#1}{#2}}

\begin{document}

\begin{titlepage}

\begin{center}

\vspace{.3in}

{\Large{\bf Relational Particle Models as Toy Models}}

\vspace{.1in}

{\Large{\bf for Quantum Gravity and Quantum Cosmology}} 

\vspace{.3in}

{\large{\bf Edward Anderson}}

\vspace{.3in}

\noindent{\em Peterhouse, Cambridge, U.K., CB21RD;}

\noindent{\em DAMTP, Centre for Mathemetical Sciences, Wilberforce Road, Cambridge, U.K., CB30WA.}

\noindent Work done while at {\em Department of Physics, Avadh Bhatia Laboratory,}

\noindent{\em University of Alberta, Edmonton, Canada, T6G 2J1.}

\end{center}

\vspace{.3in}

\begin{abstract}
 
It is argued that substantial portions of both Newtonian particle mechanics  
and general relativity can be viewed as relational (rather than absolute) theories.  
I furthermore use the relational particle models as toy models to investigate 
the problem of time in closed-universe canonical quantum general relativity. 
I consider thus in particular the internal time, semiclassical and records tentative 
resolutions of the problem of time.

\end{abstract}

\end{titlepage}

\section{Introduction}

I consider relational models \cite{BB82, Barbour, RWR, Relationalpapers, Kiefer} for the universe.  
These have two features.  
1) Temporal relationalism: that there is no physically meaningful primitive notion of time 
for the universe as a whole.
2) Spatial relationalism: that each notion of space possesses a transformation group $G$ 
which does not alter the physical content of the universe.  
1) is implemented by considering actions which are invariant under reparametrization of `label time' 
$\lambda$ by being homogeneous linear in this. 
2) is implemented by these actions being constructed out of objects natural to the 
configuration space $Q$ in question and furthermore being corrected by auxiliary variables 
corresponding to the independent infinitesimal transformations of $G$.  
Both of these implementations lead to constraints.  
In the former this is through the $n$ momenta subsequently being homogeneous of degree 0 in the 
velocities and hence functions of at most $n - 1$ independent ratios of velocities so that the momenta 
must have at least 1 relation between them (which is by definition a {\it primary constraint}).  
In the latter case, one may worry that one is giving further objective existence to $G$, since, 
by the introduction of the auxiliary variables, one is passing from an action on $Q$ to one on 
$Q \times G$.  
However, subsequent variation with respect to each of these auxiliaries produces one 
{\it secondary constraint}, which uses up {\sl two} degrees of freedom, so that one ends up on 
the quotient space $Q/G$ of equivalence classes of $Q$ under $G$ motions, so $G$ is indeed 
rendered physically irrelevant by this {\sl indirect means}.

In Sec 2, I consider $Q$ to be the space $Q(N)$ of $N$-particle positions 
and $G$ to be the group of translations and rotations, Eucl.    
This relational particle model (RPM) is a reformulation of the zero angular momentum portion of Newtonian 
mechanics, for which I furthermore provide a {\sl direct implementation} of spatial relationalism.    
In Sec 3, I explain that general relativity (GR) arises if one considers $Q$ to be the space of 
3-metrics Riem($\Sigma$) on an arbitrary spatial closed (compact without boundary) 3-geometry $\Sigma$ 
and $G$ to be the 3-diffeomorphisms on $\Sigma$, Diff($\Sigma$).  
In Sec 4, I explain that canonical quantization of GR runs into a number of difficulties concerning the 
clash between the GR and quantum notions of time in closed universes, 
and I sketch some tentative resolutions.  
In Sec 5, I use RPM's as toy models to investigate these.

\section{Relational Particle Models}

Consider a reparametrization-invariant action for particle mechanics with the Euclidean group of 
motions of flat space rendered irrelevant by passing from particle position coordinates\fn{Particle 
positions are indexed by capital letters running from 1 to $N$.  
Relative coordinates are indexed by lower-case letters running over 1 to $N - 1$.  
Spatial indices are lower-case Greek letters.  
$E$ and $V$ are the total and potential energies of each universe model. $V$ depends 
on the magnitude of relative position variables only.} $q_{\alpha}^I$ 
to $q_{\alpha}^I  - a_{\alpha}  - {\epsilon_{\alpha}}^{\beta\gamma}b_{\beta}q^I_{\gamma}$:  
\be
S[q_{\alpha I}, \dot{q}_{\alpha I}, \dot{a}_{\alpha}, \dot{b}_{\alpha}] = 2\int\d\lambda\sqrt{T(E - V)} \mbox{ , } 
T = \sum_{I = 1}^{N}m_I\delta^{\alpha\beta}
(\dot{q}_{\alpha I} - \dot{a}_{\alpha} - {\epsilon_{\alpha}}^{\sigma\rho}\dot{b}_{\sigma}q_{\rho I} )
(\dot{q}_{\beta I} - \dot{a}_{\beta} - {\epsilon_{\beta}}^{\lambda\mu}\dot{b}_{\lambda}q_{\mu I} ) 
\mbox{ } .
\ee
This is a re-interpretation of Barbour and Bertotti's work \cite{BB82}.    
Then, the momenta are
\be
p^{\alpha A} = \sqrt{\frac{E - V}{T}}m_A\delta^{\alpha\beta}
(\dot{q}_{\beta I} - \dot{a}_{\beta} - {\epsilon_{\beta}}^{\lambda\mu}\dot{b}_{\lambda}q_{\mu I} )
\mbox{ } .  
\ee
Reparametrization invariance leads to 
\be
H \equiv \sum_{I = 1}^{N}\frac{1}{2m_{I}}\delta_{\alpha\beta}p^{\alpha I}p^{\beta I} + V = E 
\mbox{ } \mbox{ (energy constraint)}
\ee
as a primary constraint via 

\noindent$\sum_{I = 1}^{N}\frac{1}{2m_{I}}\delta_{\alpha\beta}p^{\alpha I}p^{\beta I} 
= \sum_{I = 1}^{N}\frac{1}{2m_{I}}\delta_{\alpha\beta}\sqrt{\frac{E - V}{T}}
(\dot{q}_{\alpha I} - \dot{a}_{\alpha} - {\epsilon_{\alpha}}^{\sigma\rho}\dot{b}_{\sigma}q_{\rho I})
\sqrt{\frac{E - V}{T}}(\dot{q}_{\beta I} - \dot{a}_{\beta} - {\epsilon_{\beta}}^{\lambda\mu}\dot{b}_{\lambda}q_{\mu I}) 
= \frac{E - V}{T}T = E - V$, 
while $a_{\alpha}$ and $b_{\beta}$ variation lead to secondary constraints 
\be
P^{\alpha} \equiv \sum_{I = 1}^{N}p^{\alpha I} = 0  
\mbox{ } , \mbox{ } 
L^{\alpha} \equiv \sum_{I = 1}^{N} {\epsilon^{\alpha\beta}}_{\gamma} q_{\beta I}  p^{\gamma I} = 0 
\mbox{ } \mbox{ (0 total momentum and 0 total AM constraints)} \mbox{ } .
\ee 
Furthermore, this RPM is a reformulation of a portion of Newtonian mechanics, 
one restriction being $L_{\alpha} = 0$.

As elimination of $\dot{a}_{\alpha}$ and $\dot{b}_{\alpha}$ 
from the Lagrangian form of $P^{\alpha}$ and $L^{\alpha}$ is possible for this example, 
the indirectness of the above implementation of spatial relationalism (resolution of 
absolute versus relative motion debate) furthermore turns out to be unnecessary. 
Using relative Jacobi coordinates $R_{\alpha}^i$, 
which are interparticle (cluster) separations  
and which have the useful properties of automatically accounting for $P^{\alpha} = 0$ 
and preserving the form of all else in the above expressions 
(just swap $q_{\alpha}^I$ for $R_{\alpha}^i$ throughout), 
a direct implementation is (see \cite{I} for the derivation and further discussion):
\be
S(R^i_{\alpha}, \dot{R}_{\beta}^j) = 2\int\d\lambda \sqrt{(E - V)T}
\mbox{ } , \mbox{ }
T(R^i_{\alpha}, \dot{R}_{\beta}^j) = 
\sum_{i = 1}^{N - 1}\frac{M_i}{2}\dot{R}_i^2 
- \frac{1}{2}\stackrel{\B}{L}_{\alpha}(\stackrel{\B}{I}{}^{-1})^{\alpha\beta}\stackrel{\B}{L}_{\beta}
\mbox{ } , 
\ee
where $\stackrel{\B}{I}_{\alpha\beta}$ and $\stackrel{\B}{L}_{\alpha}$ are the barycentric 
inertia tensor and angular momentum respectively: 
\be
\stackrel{\B}{I}_{\alpha\beta}(R^i_{\alpha}, \dot{R}_{\beta}^j) = \sum_{i = 1}^{N - 1}M_i
\left(
|R_i|^2\delta_{\alpha\beta} - R_{\alpha i}R_{\beta i}  
\right)
\mbox{ } \mbox{ and } \mbox{ } 
\stackrel{\B}{L}_{\alpha}(R^i_{\alpha}, \dot{R}_{\beta}^j) = 
{\epsilon_{\alpha}}^{\beta\gamma}\sum_{i = 1}^{N - 1}M_iR_{\beta i}\dot{R}_{\gamma i}\mbox{ } . 
\ee

\section{GR as a Relational Theory}

A slight re-interpretation \cite{RWR} of the Baierlein--Sharp--Wheeler \cite{BSW} action is a relational 
formulation for spatially compact without boundary GR\fn{Here, $\gamma_{\alpha\beta}$ is the spatial 
3-metric with determinant $\gamma$, covariant derivative $D_{\alpha}$ and Ricci scalar $R$. $\Lambda$ is 
the cosmological constant.}  
\be
S = \int \textrm{d}t \int \d^3x \sqrt{\gamma} 
\sqrt{(\Lambda + R)T_{GR}    }
\mbox{ } , \mbox{ }
T_{GR} = 
(\gamma^{\alpha\gamma}\gamma^{\beta\delta} - \gamma^{\alpha\beta}\gamma^{\gamma\delta})
(\dot{\gamma}_{\alpha\beta} - \pounds_{\dot{s}}\gamma_{\alpha\beta})
(\dot{\gamma}_{\gamma\delta} - \pounds_{\dot{s}}\gamma_{\gamma\delta})
\label{BSW} \mbox{ } .  
\ee
For, note that this is reparametrization-invariant and is built using not $\gamma_{\alpha\beta}$ 
but $\gamma_{\alpha\beta} - \pounds_s\gamma_{\alpha\beta}$ where $\pounds_{s}$ is 
the Lie derivative with respect to the 3-diffeomorphism auxiliary $s_i$.  
(In fact, this action emerges as one of only a few consistent options upon considering far more general 
actions built from these relational first principles on the configuration space of 3-metrics on a 
compact without boundary spatial 3-manifold \cite{RWR}.)
Now, the gravitational momenta are 

\noindent$\pi^{\gamma\delta} = \sqrt{\frac{\Lambda + R}{T}}
(\gamma^{\alpha\gamma}\gamma^{\beta\delta} - \gamma^{\alpha\beta}\gamma^{\gamma\delta})
(\dot{\gamma}_{\alpha\beta} - \pounds_{\dot{s}}\gamma_{\alpha\beta})$.  
Then 
\be
{\cal H}  \equiv \frac{1}{\sqrt{\gamma}}
\left(
\gamma_{\alpha\gamma}\gamma_{\beta\delta} - \frac{1}{2}\gamma_{\alpha\beta}\gamma_{\gamma\delta}
\right)
\pi^{\alpha\beta} \pi^{\gamma\delta}  -  \sqrt{\gamma}R  = 0 \mbox{ } \mbox{ (Hamiltonian constraint) } 
\label{Ham}
\ee
follows as a primary constraint by a working closely related to that displayed in the previous section, 
and 
\be
{\cal H}_{\alpha}  \equiv -2D_{\beta}{\pi_{\alpha}}^{\beta} = 0              
\mbox{ } \mbox{ (momentum constraint) }
\label{Mom} 
\ee
follows from $s_{\alpha}$-variation.

Note the close parallels between this and the previous section: energy $E$ and cosmological constant 
$\Lambda$ play the same role, both actions are of square root form, leading to 
quadratic constraints ($H$ and ${\cal H}$ respectively), and in each case variation 
with respect to auxiliaries leads to linear constraints ($P_{\alpha}$, $L_{\alpha}$, and 
${\cal H}_{\alpha}$, respectively).   
Elimination from the Lagrangian form of the linear constraints is a significant procedure 
in both cases: in Sec 2 it provides an explicit direct resolution of 
the absolute versus relative motion debate, while in Sec 3 it now constitutes the 
well-known {\sl thin sandwich conjecture}.  
The relative configuration space quotient of the $Q(N)/\mbox{Eucl}$
parallels the superspace \cite{DeWitt} quotient $\mbox{Riem}(\Sigma)/\mbox{Diff}(\Sigma)$ 
in being curved and stratified.  
Furthermore, one can then attempt relative configuration space quantization 
much as one can attempt superspace quantization.

\section{Problems with Time and Closed Universes in GR}

My interest in the above similarities stems from conceptual and technical problems which 
one encounters in canonical GR with the quantum form of ${\cal H} = 0$ \cite{Kuchar92, Isham93, 
Barbour} -- in the configuration representation, this gives what is prima facie a 
time-independent Schr\"{o}dinger equation $\hat{{\cal H}}\Psi = 0$ rather 
than one which is dependent on some notion of time, $\tau$: 
$\hat{{\cal H}}\Psi = i\hbar\frac{\delta \Psi}{\delta \tau}$.  
I hope that light will be shed on the conceptual viability of various suggested resolutions of this
by considering the RPMs' analogous yet technically simpler quantum equation $(H - E)\Psi = 0$.

One of these resolutions  is \cite{Kuchar92, Isham93} that there is really a time hidden within 
${\cal H}$ itself.    
This is based on the hope that there exists a canonical transformation which separates out four 
embedding variables and two true degrees of freedom of GR from the six 3-metric variables.   
This would produce ${\cal H}^{true} =$ (a linear combination of embedding momenta), which clearly yields 
a time-dependent Schr\"{o}dinger equation upon quantizing in the new configuration representation.  
The York time approach (see e.g. \cite{Isham93}) is one such attempt.  
Ignoring the solution of the momentum constraint for simplicity of presentation 
(so that the below involves one time function rather than four embedding functions) 
this works as follows.  
A canonical transformation permits the York time $\tau_{Y} 
= \frac{2}{3\sqrt{\gamma}}(\gamma_{\alpha\beta}\pi^{\alpha\beta})$ 
to serve as a coordinate while its conjugate quantity $\sqrt{\gamma}$ is now a momentum.  
Then the Hamiltonian constraint is replaced by 
$-{\cal H}^{true} = \sqrt{\gamma} = \chi^6$,  
for $\chi$ the solution of the conformally-transformed ${\cal H}$ 
\be
8\nabla^2\chi = \frac{\pi^2}{6}\chi^5 + R\chi - 
\pi_{\alpha\beta}\pi^{\alpha\beta}\chi^{-7} 
\mbox{ } \mbox{ (Lichnerowicz--York equation) } \mbox{ } .   
\label{LYE}  
\ee 
Then quantization gives 
\be
i\hbar\frac{\delta\Psi}{\delta \tau_{Y}   } 
= \widehat{    {\cal H}^{true}    }\Psi 
\mbox{ } .  
\ee
The obstruction to this particular resolution is that how 
to solve the complicated quasilinear elliptic equation (\ref{LYE}) is not in practice generally known, 
so the functional dependence of ${\cal H}^{true}$ on the other variables is not known, so the  
quantum `true Hamiltonian' $\widehat{    {\cal H}^{true}    }$ cannot be explicitly defined.  

Other resolutions consider time not to exist fundamentally, 
but rather to be an {\sl emergent} concept.  
I consider two such approaches to quantum gravity: 
the semiclassical approach \cite{Kiefer} and the records approach \cite{Barbour, H99}.  
Both have been associated with splits into heavy and light modes, $h_{A^{\prime}}$ and 
$l_{A^{\prime\prime}}$ respectively.   
The semiclassical approach then uses the WKB ansatz for the wavefunction of the universe: 
\be
\Psi = e^{    iM_{h}W(h_A^{\prime})/\hbar    }
\psi(h_{A^{\prime}}, l_{A^{\prime\prime}})
\mbox{ } \mbox{ } \mbox{ (WKB ansatz for the wavefunction) } , 
\ee
where $W$ is the principal function.  
Substituting this into $\hat{\cal H}\Psi = 0$, 
one can peel off the Hamilton--Jacobi equation as the leading order terms, 
and moreover keeps the derivative cross term to the next order of approximation.  
It is this that supplies the emergent WKB time $\tau_{WKB}$.  
In the case of `heavy gravitational modes' supplying WKB time to 
`light minimally coupled matter modes' $\phi$ 
(which contribute additive portions ${\cal H}^{\phi}$ and 
${\cal H}_{\alpha}^{\phi}$ to ${\cal H}$ and ${\cal H}_{\alpha}$ respectively), 
one has\fn{Here, $\alpha$ is a lapse (proper time elapsed) e.g. emergent 
from the reparametrization-invariant form of the action (\ref{BSW}).  
The shift $\beta_{\alpha}$ is the same notion as $\dot{s}_{\alpha}$.} 
\be
i\hbar\frac{\delta\psi}{\delta \tau_{WKB}} = 
\left(
\int \d^3x(\alpha{\cal H}^{\phi} + \beta^{\alpha}{\cal H}_{\alpha}^{\phi})
\right)
\psi 
\mbox{ } \mbox{ } \mbox{ (emergent WKB time-dependent Schr\"{o}dinger equation of GR) }
\label{WKBTISEGR}
\ee
Records approaches treat particular sorts of instantaneous configurations as primary 
and then attempt to reconstruct a semblance of dynamics/history from these.  
The main problems of these schemes are 
justifying the WKB ansatz and not being fully worked out respectively.

\section{Investigation using RPM's as Toy Models}

Via a carefully-ordered approach using Jacobi coordinates, a range of simple RPM's 
can be treated in good part using the usual mathematics employed in QM \cite{I}.  
This means that, at the simplest level, absolutism has not misled the conceptual development of QM.  
This range of simple RPM's also permits the study of some simple 
features of closed universes: truncations and gaps in the eigenspectrum  
(which can go away with increase in complexity), 
and the `limited resource' effect of a fixed and finite energy for the whole universe and 
subsystem angular momentum balance effects (which do not go away with increase in complexity  
but would seem likely to evade notice in large universes in which only small subsystems 
are ever studied in practice).

Further investigation \cite{II} reveals that these RPM's 
have an `Euler' internal time  
\be
\tau_{\cal E} \equiv \sum_{I = 1}^{N}q_{I\alpha}  p^{I\alpha} \mbox{ } .
\ee
For this to be a time, it is important that this is monotonic;  
this follows for a number of substantial cases from the Lagrange--Jacobi identity: 
\be
\dot{\tau}_{\cal E} = 2T + kV = 2E + (k - 2)V \mbox{ } , 
\ee
for $V$ is homogeneous of degree $- k$.  
$\tau_{{\cal E}}$ is conjugate to the scale variable $\sigma$ (which is the 
logarithm of a quantity proportional to the square root of the moment of inertia of the system)
the rest of the variables are shapes and their momenta.  
A canonical transformation can then be applied so that $\tau_{\cal E}$ indeed 
becomes a coordinate and $\sigma$ its momentum.  
The idea is then that $H = E$ is to be interpreted as equation for $\sigma$. 
I have done this e.g for simple 3-particle models in 1-d in \cite{II}.   
Then, the scale variable is $\sigma = $ln$\sqrt{R_1^2 + R_2^2}$ 
and there is one shape variable, which may be chosen to be of the form $S = \mbox{Arctan}(R_1/R_2)$. 
Then 
\be
E \equiv H(\tau_{\cal E}, S; -\sigma, {P}_S) 
= \frac{e^{-2\sigma}}{2\mu}(\tau_{\cal E}^2 + P_S^2) + V(\sigma, S) \mbox{ } .
\ee
For a number of standard potentials, this is explicitly soluble as 
$\sigma = \sigma(\tau_{\cal E}, S, P_S) \equiv - H^{true}(\tau_{\cal E}, S, P_S)$
For each of these one may then pass to an explicit configuration representation Euler internal time 
dependent Schr\"{o}dinger equation 
\be
i\hbar\frac{\pa\Psi}{\pa\tau_{{\cal E}}} = \widehat{H^{true}}
(\tau_{\cal E}, S, \hat{P}_S)\Psi
\ee
Thus, after stripping Newtonian mechanics bare of absolute space and absolute time, 
I find that some portions of it nevertheless have an internal time hidden within!  
Having got round the not explicitly constructible impasse which plagues GR, 
there is some value in investigating next whether these internal time dependent Schr\"{o}dinger 
equations are quantum mechanically well-defined and what properties their solutions have, 
so as to infer how sound internal time approaches are.

As regards the emergent time resolutions, I begin by setting up the heavy--light split for the 
relative Jacobi coordinates $R_{\alpha}^i$ of the RPM's.  
Consider RPM's for which the $R_{\alpha}^i$ subdivide into heavy $h_{\alpha}^{i^{\prime}}$ and light 
$l_{\alpha}^{i^{\prime\prime}}$ coordinates according to the magnitudes 
of the associated masses (which are the reduced masses of clusters of the 
original particle-position masses) being such that $M_{h} >> M_{l}$.   
This is possible e.g. for two $h$ particles of similar mass $M$ and one $l$ particle of mass $m$, 
whence there is then one heavy Jacobi coordinate $\tilde{h}_{\alpha}$ and one light one 
$\tilde{l}_{\alpha}$ with 
\be
M_{\tilde{h}} \approx M/2 >> m = M_{\tilde{l}} \mbox{ } .
\label{Po}
\ee

The semiclassical approach then involves the WKB ansatz
\be
\Psi = e^{iM_{\tilde{h}}W(\tilde{h}_{\alpha})/\hbar}\psi(\tilde{h}_{\alpha}, \tilde{l}_{\alpha})
\mbox{ } .  
\label{WKB}
\ee  
Substitute this into $(\hat{H} - E)\Psi = 0$ and $\hat{L}\Psi =0$,  
keeping the cross-term proportional to 
$\frac{\pa W}{\pa \tilde{h}_{\alpha}}\frac{\pa \psi}{\pa \tilde{h}^{\alpha}}$.  
Then at zeroth order the corresponding Hamilton--Jacobi equation appears, while
to first order\fn{Here $\dot{a}$ is a `lapse' emerging 
from the reparametrization-invariant action used. $\dot{b}^{\alpha}$ is as in Sec 2.}
\be
i\hbar\frac{\pa\psi}{\pa \tau_{WKB}} = 
\left(
\dot{a}\hat{H}^{(l-part)} - \dot{b}^{\alpha}\hat{L}_{\alpha}^{(l-part)}
\right)
\psi
\ee
eventually arises \cite{II}, 
which is a simple analogue of the emergent WKB time-dependent Schr\"{o}dinger 
equation of GR (\ref{WKBTISEGR}).  
While I addressed a number of objections \cite{BS} specific to the semiclassical approach of RPM's 
in \cite{I}, a remaining objection {\sl in general} to the semiclassical approach is justifying 
the WKB-type ansatz in the context of whole universes.  
I have not resolved this foundational issue,  
but hope that RPM's will be a useful arena in which to investigate whether this can be justified.

A records approach built on the above $h$--$l$ set-up follows along the lines of 
\cite{H99} which considers one $h$ particle moving through a medium of $l$ particles which it 
disturbs into motion.  
Subsequent instants consist of the particles' positions and momenta.  
It is these instants which are the records, and the motion or history of the large particle 
can then be reconstructed (perhaps to some approximation) from them.   
This approach has the complicating feature that a potential including $h$--$l$ coupling is required, 
and the simplifying feature that the environment of $l$ particles need not be populous.   
This notion of record can be adapted to e.g. a 1-d 3-particle RPM as follows.    
Consider the 2 $h$ and 1 $l$ particle situation of (\ref{Po}).  
In relational terms, this situation is the motion of a $\tilde{h}$ and a $\tilde{l}$ interparticle 
(cluster) separations (Jacobi coordinates).  
If these have coupled potentials so that the $\tilde{h}$ separation disturbs the $\tilde{l}$ separation 
into motion, subsequent record-instants consist of inter-particle (cluster) 
separations and their momenta with respect to label time.  
An additional issue to investigate is: what in nature causes the selection of records rather than 
instants from which a semblance of dynamics/past history cannot be reconstructed.  
Barbour's approach \cite{Barbour} speculates that the asymmetry of the 
underlying curved stratified quotient configuration space causes concentration of the 
wavefunction on records rather than other instants.  
As RPM's share these features, they 
may serve to investigate this possibility.

Investigating the schemes of the last two paragraphs in detailed particular examples 
involves further work of greater complexity than in my current work \cite{II}. 
RPM's may then be a promising arena to investigate, 
in cases in which two or more of the above resolutions of the problem of time 
exist, whether these are identical, approximate, or entirely distinct resolutions.

\mbox{ } 

\noindent{\bf Acknowledgments}

\mbox{ } 

\noindent Don Page, Julian Barbour, Niall \'{O} Murchadha and Gary Gibbons for discussions, Claire 
Bordenave for accommodation and help with the computer, the Killam Foundation and then Peterhouse for funding.


%

\end{document}